\documentclass[twocolumn,floatfix]{revtex4-1}

\usepackage{amsmath}
\usepackage{graphicx}
\usepackage{amssymb}
\usepackage{subfigure}
\usepackage{float}
\usepackage{placeins}

\begin{document}



\title{Impact of anisotropy on vortex clusters and their dynamics}

\author{J. Stockhofe}
\author{S.\ Middelkamp}
\author{P.\ Schmelcher}
\affiliation{ Zentrum f\"ur Optische Quantentechnologien, Universit\"at
Hamburg, Luruper Chaussee 149, 22761 Hamburg, Germany
}
\author{P.G.\ Kevrekidis}
\affiliation{Department of Mathematics and Statistics, University of Massachusetts,
Amherst MA 01003-4515, USA}

\begin{abstract}
We investigate the effects of anisotropy on the stability and dynamics of vortex cluster states which arise in Bose-Einstein condensates. 
Sufficiently strong anisotropies are shown to stabilize states with arbitrary numbers of vortices that are highly unstable in the isotropic limit. 
Conversely, anisotropy can be used to destabilize states which are stable in the isotropic limit. 
Near the linear limit, we identify the bifurcations of vortex states including their emergence from linear eigenstates, 
while in the strongly nonlinear limit, a particle-like description of the dynamics of the vortices in the anisotropic trap is developed. 
Both are in very good agreement with numerical results.
Collective modes of stabilized many vortex cluster states are demonstrated.
\end{abstract}

\maketitle

{\it Introduction.} 
Vortices constitute one of the nonlinear wave structures that have
received considerable attention not only within Bose-Einstein condensates
(BECs)
in dilute atomic gases \cite{stringari,pethick,emergent,fetter0,review},
but also in nonlinear optics ~\cite{yuripismen,dragomir,Kivshar-LutherDavies,YSKPiO}, among other fields~\cite{Pismen}. Nevertheless, admittedly, BECs
constitute one of the pristine settings where structural and
dynamical properties of single- and multi-vortex or multi-charged-vortex
states can be investigated experimentally and compared to theoretical
predictions. It is for that reason that multiple techniques
were developed to produce such vortices through 
phase-imprinting \cite{Matthews99}, stirring \cite{Madison00} or
nonlinear interference \cite{BPAPRL} and few vortex states \cite{Madison01},
vortex lattices \cite{Raman} and higher charge states \cite{S2Ket} were
generated experimentally.

Although single- and multi-charge vortices were intensely studied, one of the themes that has
received somewhat less attention is that of clusters of few vortices.
Central questions here include the possible geometries and structures of vortex clusters as well as their stability.
Furthermore, the dynamics of cluster configurations is of immediate interest:
due to the intricate interaction among the vortices, we indeed expect a rich ``vibrational'' dynamics of vortex clusters.
In the literature mostly the vortex dipole has been considered 
\cite{crasovan,mott1,mott2,komineas,pgk1} and it has been argued to
be a robust dynamical configuration, while other states such as
tripoles, quadrupoles, etc. are generally considered to be fragile \cite{mott2,pgk1}. 
There are interesting connections between these states
and dark solitonic stripes \cite{pantoflas,komineas,pgk1}, since some of the multi-vortex states arise from the instabilities
of such dark-soliton stripes. 
The current interest in vortex clusters is also stimulated by several very recent experiments on the preparation of vortex dipoles \cite{BPA_recent,dshall_recent}
and three-vortex states~\cite{bagnato} as well as their dynamics.

Our aim in this work is to present a detailed discussion of the existence,
stability and dynamical properties of aligned vortex states by encompassing anisotropy i.e., we
include the isotropic case  of the above works as a 
particular limit and extend it to arbitrary anisotropies. 
In the case of co-rotating
vortices (in BECs with rotation), works in a similar spirit are those of
\cite{busch}, although in the latter case the multi-vortex state may
indeed be the ground state of the system. Here we focus on the anisotropic
variants of the cases
relevant to the above recent experiments.
We employ two complementary theoretical methods:
chiefly and for large atom numbers, 
we extend the particle picture for vortices, developed for isotropic settings in \cite{pgk1}, to the anisotropic regime.
The resulting set of coupled ordinary differential equations (ODEs) is then used to gain insight into the stability of stationary multi-vortex 
configurations.
On the other hand, we report on near-linear techniques valid in the 
limit of small atom numbers which allow us to make a very general prediction 
about the stabilization of a class of vortex states. 
These complementary theoretical approaches are corroborated 
by the explicit numerical solutions of the corresponding 2D Gross-Pitaevskii equation (GPE).
We identify the vortex clusters via fixed point methods for a
wide range of atom numbers and strengths of the anisotropy. 
Subsequently a Bogoliubov-de Gennes (BdG) analysis of these states is performed, thereby yielding information on their 
linear stability.
To examine their robustness beyond linear stability theory, we disturb vortex states with white noise and employ real-time propagation.
Our key conclusion is that anisotropy can controllably stabilize
a whole class of unstable states of the isotropic limit, but it can also controllably
destabilize stable states of that limit; these predictions should be 
directly testable in the experimental settings of refs. \cite{BPA_recent,dshall_recent,bagnato}.

{\it Setup.} The general framework is set by the dimensionless GPE:
\begin{equation}
 {\rm i} \partial_t \psi(x,y,t) = \left[ -\frac{1}{2} \Delta + V(x,y) + | \psi(x,y,t) |^2 \right] \psi(x,y,t).
\end{equation}
In the BEC setting this implies that length, time, energy and density $|\psi|^2$ are measured in units of $a_z$, $\omega_z^{-1}$, $\hbar \omega_z$ and $(2\sqrt{2\pi}aa_z)^{-1}$, respectively; $a$ is the s-wave scattering length. $\omega_z$ and $a_z$ refer to the oscillator frequency and 
length in the z-direction. The anisotropic trap is $V(x,y)=(\omega_x^2 x^2 + \omega_y^2 y^2)/2$. Stationary vortex states will be sought in the form
$\psi(x,y,t)=\exp(-{\rm i} \mu t) u(x,y)$, where $\mu$ denotes the chemical
potential.

\begin{figure}[ht]
\centering
\includegraphics[width=8cm]{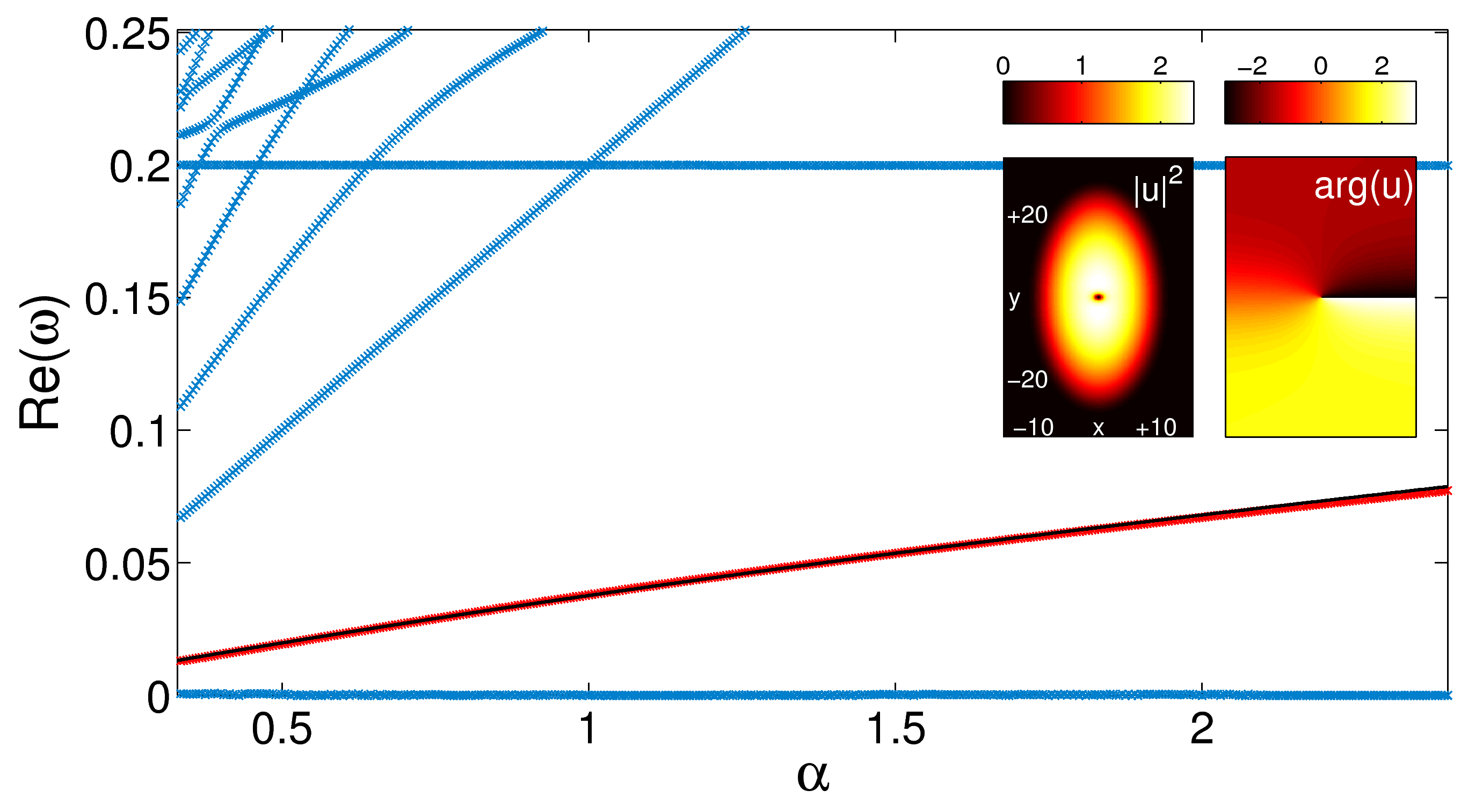}
\caption{BdG spectrum of the single vortex as a function of the anisotropy parameter $\alpha$ at $\mu=2.5$.
The prediction for the precession frequency obtained from the particle picture is shown as solid black line which is almost completely covered by the corresponding numerical data. 
The insets show contour plots of the vortex state's square modulus and phase profiles at $\alpha=0.4$, $\mu=2.5$.}
\label{figvor}
\end{figure}

\begin{figure}[ht]
\centering
\includegraphics[width=8cm]{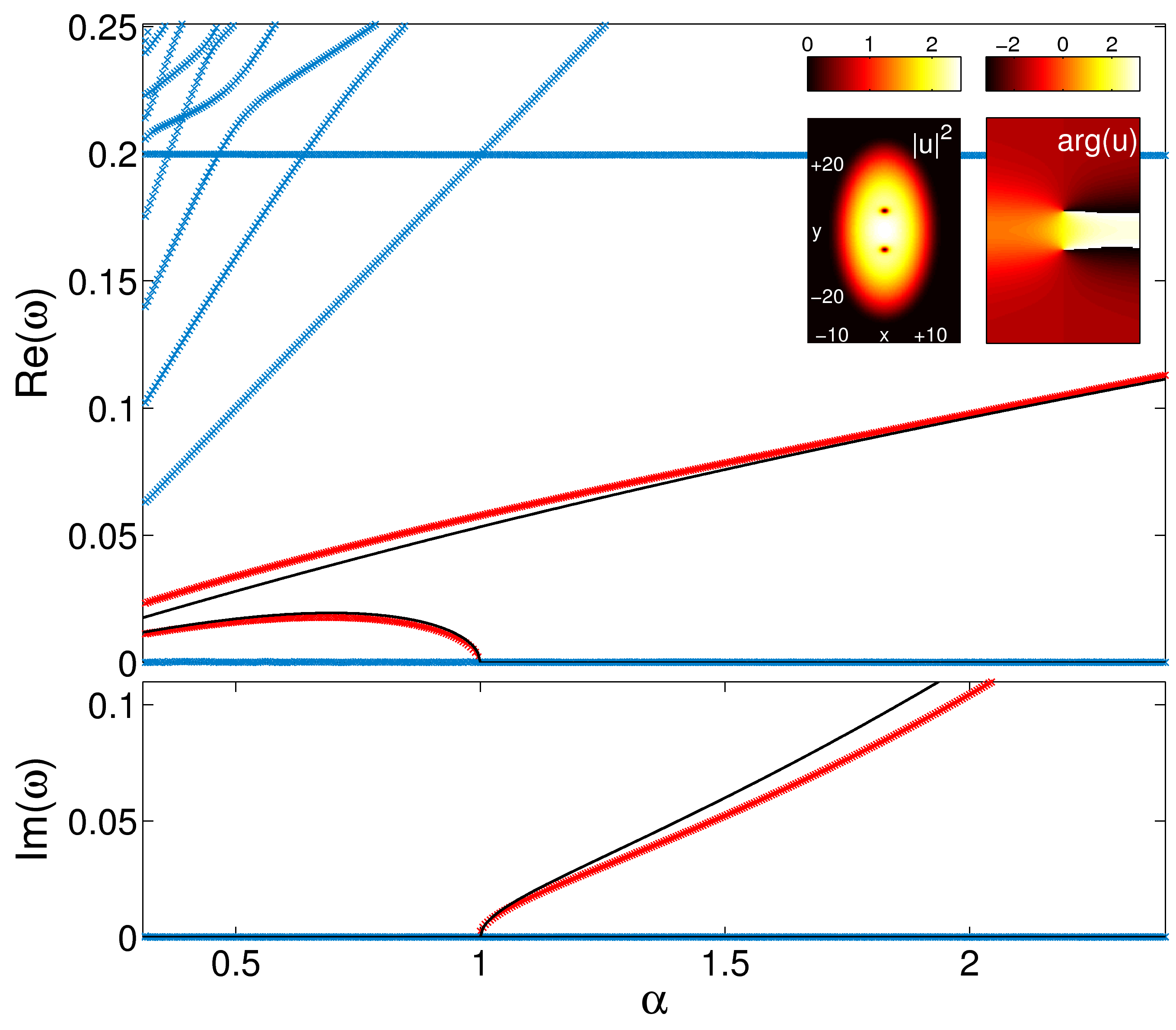}
\caption{BdG spectrum of the vortex dipole as a function of the anisotropy parameter 
$\alpha$ at $\mu=2.5$; predictions from the particle picture are shown as black lines. 
The inset shows the dipole at $\alpha=0.4$, $\mu=2.5$.}
\label{figvd}
\end{figure}

\begin{figure}[ht]
\centering
\includegraphics[width=8cm]{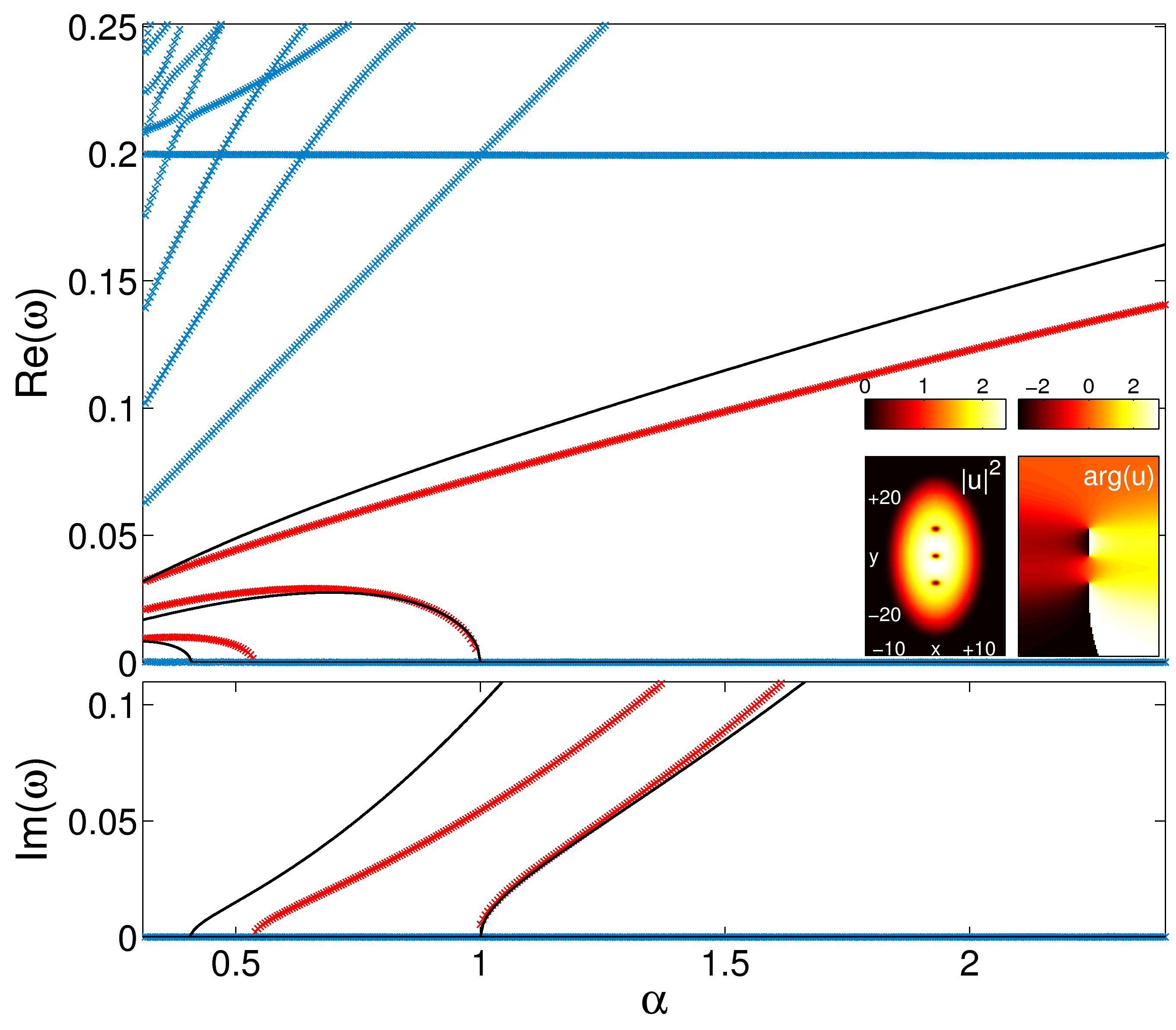}
\caption[Optional caption for list of figures]{\label{fig:bdg_vt} BdG 
spectrum of the vortex tripole as a function of the anisotropy parameter $\alpha$ at $\mu=2.5$; predictions from the particle picture shown as black lines. The inset 
shows the tripole state at $\alpha=0.4$, $\mu=2.5$.}
\end{figure}

{\it Results.} The single vortex, dipole and tripole states will be the prototypical ones
considered here. They
are shown as insets in Figs. \ref{figvor}, \ref{figvd} and 
\ref{fig:bdg_vt}, respectively, 
for an anisotropic harmonic potential (with $\omega_x=0.2$, $\omega_y=0.08$) 
and at $\mu=2.5$.
We refer to these as ``aligned vortex states''.
While the single vortex and the vortex dipole are stable in the isotropic 
limit, all higher aligned vortex states (tripole, quadrupole,
quintopole, etc. \cite{pgk1}) are subject to dynamical instabilities in
that limit.

The vortex particle picture from \cite{pgk1,JPB} can be naturally extended to 
take anisotropic confinement into account.
In the asymptotic regime of high values of $\mu$, one can use the work of \cite{fetter} to examine the consequences of anisotropy on the vortex precession.
This yields the following equations of motion for a single vortex:
$\dot{x} = - \omega_y^2 Q y$ and $\dot{y} = \omega_x^2 Q x$, where $Q=\ln(A \mu/\omega_{\text{eff}})/(2 \mu)$, $A \approx 2 \sqrt{2} \pi$ is a numerical 
constant \cite{pgk1} and 
$\omega_{\text{eff}}=\sqrt{(\omega_x^2+\omega_y^2)/2}$. 
This implies that the precession frequency in the anisotropic trap is given by $\omega_{\text{pr}}=\omega_x \omega_y Q$, which yields the 
isotropic limit of \cite{pgk1,JPB} 
when $\omega_x=\omega_y$.
Our numerical results on the single vortex state are in excellent agreement with the predictions from the above particle picture:
At high enough values of $\mu$, for every 2D aspect ratio 
$\alpha=\omega_y/\omega_x$ that we have considered, the stationary single vortex solution exists, is dynamically stable and the value of
the single 
anomalous mode in its BdG spectrum (responsible for vortex precession 
\cite{JPB}) 
 is accurately described by $\omega_{\text{pr}}$ above, as shown in
Fig. \ref{figvor}.
Note that in our numerical calculations we kept $\omega_x=0.2$ fixed and varied $\omega_y$ to scan different values of $\alpha$.
Notice also that, in addition to the vortex internal modes, in both this and in following BdG plots, the BdG analysis captures the modes of the background state on top of which the vortex ``lives''.
These include e.g. the dipolar frequency at $\omega=0.2$ due to $\omega_x=0.2$, the dipolar frequency of oscillation around the $y$-axis (which is $0.2$ in the isotropic case, but varies as $\omega_y$ varies), as well as higher order such as quadrupolar modes.

If we now consider multiple vortices, then the corresponding large
density particle picture should include the combined effects of the 
anisotropic precession and vortex interaction dynamics. Thus, the equations of 
motion of a vortex-antivortex pair i.e., a vortex dipole, will read:
\begin{eqnarray}
\dot{x}_1 &=& - \omega_y^2 Q y_1 + B \frac{y_1-y_2}{2 \rho^2}
\label{eqn3}
\\
\dot{y}_1 &=& \omega_x^2 Q x_1 - B \frac{x_1-x_2}{2 \rho^2}
\label{eqn4}
\\
\dot{x}_2 &=&  \omega_y^2 Q y_2 - B \frac{y_2-y_1}{2 \rho^2}
\label{eqn5}
\\
\dot{y}_2 &=& -\omega_x^2 Q x_2 + B \frac{x_2-x_1}{2 \rho^2},
\label{eqn6}
\end{eqnarray}
where $\rho^2=(x_1-x_2)^2+(y_1-y_2)^2$ and $B \approx 1.95$.
These equations suggest that the two effects acting on the vortices can balance each other in which case an equilibrium position is obtained, described here by $x_{1,2}=0$ and $y_{1,2}=\pm \sqrt{B/(4 \omega_y^2 Q)}$. 
One can also predict the linearized dynamics around this position,
finding two pairs of frequencies:
$\omega_{1,2} = \pm \sqrt{2}  \omega_{\text{pr}}$
and
$\omega_{3,4} = \pm \omega_{\text{pr}} \sqrt{1-\alpha^2}$,
where $\omega_{\text{pr}}$ is the precession frequency 
of a single vortex in the anisotropic trap.
These two ``anomalous modes'' of the two-vortex state are shown in 
Fig. \ref{figvd}, together with the numerically calculated BdG spectrum of 
the vortex dipole. 

Here and in Figs. \ref{figvor} and \ref{fig:bdg_vt}, 
the coloring of the BdG modes is chosen as follows:
in the real part of the spectrum, the anomalous modes (of negative energy)
corresponding to intrinsic vortex motions
are shown in red (gray), while the other modes which are modes of the background condensate are denoted in blue (light gray).
The predictions from the linearization of the particle picture ODEs are shown in black (dark). Notice again the excellent agreement between the theoretical
prediction for the frequencies of the internal modes of the vortex pair
and the corresponding numerical results.
For any $\alpha \neq 1$ the rotational symmetry 
is broken and the corresponding mode in the BdG spectrum (which is present for any stationary solution in isotropic settings) deviates from its formerly
vanishing eigenfrequency.
Remarkably, for $\alpha>1$ this mode becomes purely imaginary, signaling destabilization of the dipole (and further destabilization of any higher 
order) configuration.
This is in agreement with the ODE linearization results and
with physical intuition suggesting that anisotropy aligning the BEC
with the vortex cluster should enhance its stability while anisotropy in
the opposite direction should destabilize it.

This particle approach can be generalized to any of the aligned vortex states.
The case of three vortices (``vortex tripole'') located in equilibrium at $x_{1,2,3}=0$ and $y_2=0$, while $y_1=-y_3=\sqrt{B/(4 \omega_y^2 Q)}$ leads naturally
to three anomalous mode pairs (the number of
such modes associated with vortex motions is equal to the number of
vortices in the cluster) with frequencies:
\begin{eqnarray}
\omega_{1,2} &=& \pm \sqrt{2}  \omega_{\text{pr}} \sqrt{1-\alpha^2}
\label{eqn9}
\\
\omega_{3,4,5,6} &=& \pm \omega_{\text{pr}} \sqrt{4-5 \alpha^2 \pm \sqrt{9+2\alpha^2 +25 \alpha^4}}.
\label{eqn10}
\end{eqnarray}
In the isotropic limit of $\alpha=1$, these yield an 
unstable mode at ${\rm i} \sqrt{7} \omega_{\text{pr}}$, a neutral one at $0$ (due to rotational invariance) and a precession frequency of 
$\sqrt{5} \omega_{\text{pr}}$.
Concerning stability, the most important conclusion here is that there is 
a {\it critical anisotropy} of $\alpha=1/\sqrt{6} \approx 0.408$, below which 
the tripole becomes completely stabilised. I.e., not only can 
anisotropy (in the ``wrong direction'' i.e., $\alpha>1$) destabilize
stable vortex clusters such as the dipole, but also in the ``right 
direction'' (i.e., $\alpha<1$), it can stabilize not only the tripole, but in 
fact any higher order aligned vortex state (e.g. we have found stabilization in an aligned quadrupole not shown here). 
What changes is the precise critical value of $\alpha$ below which this stabilization occurs.
Examining the numerically found BdG spectrum of the tripole 
shown in Fig. \ref{fig:bdg_vt}, we can see that the particle picture still captures the overall behaviour of the relevant modes.
Most importantly, stabilization of the tripole for small values of $\alpha$ is correctly predicted by the particle picture.
The critical value of $\alpha$ obtained from the BdG analysis is given by $\alpha \approx 0.5$, slightly deviating from the predicted critical anisotropy from the ODE approach.

Stabilization of the aligned vortex states for small $\alpha$ can also be understood from the (near-linear) 
limit of the number of atoms $N \rightarrow 0$, in which the GPE reduces 
to the linear Schr{\"o}dinger equation with the chemical potential 
$\mu$ playing the role of energy.
In this regime, few-mode Galerkin-type methods have been developed in 
\cite{galerkin,pgk1} 
for the isotropic case which can be extended to arbitrary $\alpha$.
More specifically,
for small particle numbers the aligned vortex states and the soliton stripe 
states from which they bifurcate as the parameter $\mu$ is increased are well 
described by superpositions of harmonic oscillator eigenstates \cite{pgk1}.
This picture is also applicable in anisotropic settings, and it is found that 
changing $\alpha$ (and thus shifting the energies of the linear eigenstates) 
qualitatively affects the bifurcation diagram and, in turn, the stability 
of the states involved.

In particular, the soliton stripe state with zero density along (and a phase jump of $\pi$ across) the $y$-axis can be traced back to the harmonic oscillator state $\psi_{10}(x,y)=\psi_1(x) \psi_0(y)$ in the linear limit, where
$\psi_n$ denotes the $n$-th excited state of the one-dimensional
quantum harmonic oscillator.
As $\mu$ is increased, the aligned vortex 
states subsequently bifurcate from this solitonic branch as has been 
shown for the isotropic case in Ref. \cite{pgk1}.
For $\alpha = 1$, except for the single vortex state all aligned vortex states 
bifurcate from the soliton stripe branch at finite values of the particle number.
Since aligned $n$-vortex states emerge from the soliton stripe due to an 
admixture (to $\psi_{10}$) of $\pm \rm{i} \psi_{0n}$ oscillator components, 
and admixtures with high $n$ are energetically suppressed, they occur at 
correspondingly high values of $\mu$.
In the isotropic case, the first vortex state bifurcating from the solitonic branch at finite $N$ is the vortex dipole, then the tripole, quadrupole and so on.
The dipole inherits the soliton's stability, while the soliton itself is destabilized due to this bifurcation.
The tripole, quadrupole etc. then bifurcate from the already unstable solitonic branch, and thus are unstable themselves.

In the following, let us apply the Galerkin-type approach (see details
in \cite{pgk1,galerkin})
to describe the bifurcations leading to the lowest aligned vortex states in the presence of anisotropy.
The critical value of the particle number and the chemical potential, respectively, at which the aligned $n$-vortex state is predicted to bifurcate from the soliton stripe branch are given by \cite{pgk1}
\begin{eqnarray}
 N_{\text{cr}} &=& \frac{\omega_{0n}-\omega_{10}}{A-B}
\label{eqncr}
\\
 \mu_{\text{cr}} &=& \omega_{10} + A N_{\text{cr}},
 \label{eqmucr}
\end{eqnarray}
where $A=\int \psi_{10}^4 dxdy$, $B=\int \psi_{10}^2 \psi_{0n}^2 dxdy$, and $\omega_{10} = \frac{3}{2}\omega_x + \frac{1}{2}\omega_y$, $\omega_{0n}=\frac{1}{2}\omega_x + (n+\frac{1}{2})\omega_y$ denote the energies of $\psi_{10}$ and $\psi_{0n}$, respectively.
Note that in all cases relevant to our discussion $A > B$.

\begin{figure}[ht]
\centering
\subfigure[$\alpha=1.5$]{
\includegraphics[height=4.8cm]{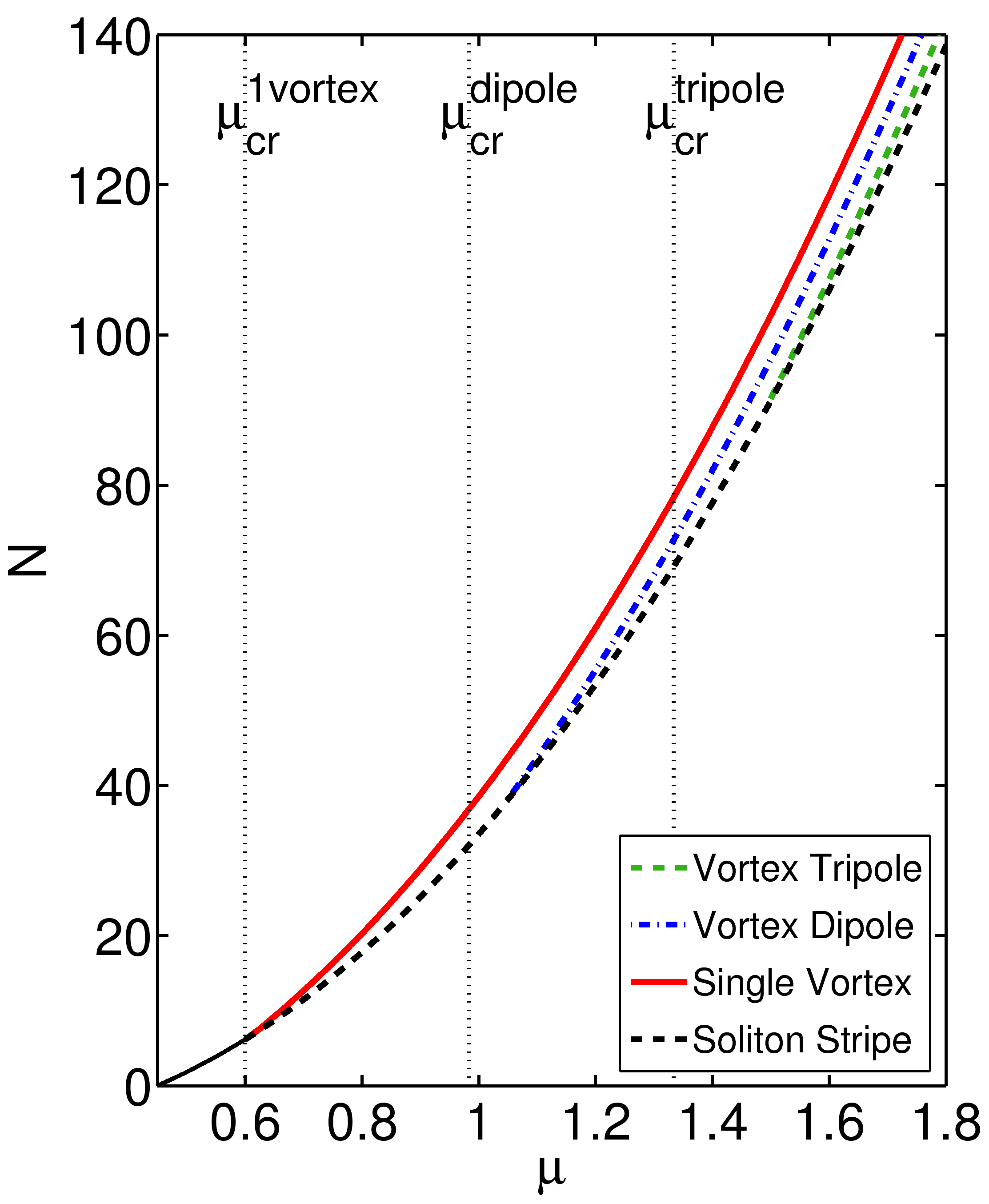}
\label{fig:bif1-5}
}
\subfigure[$\alpha=0.7$]{
\includegraphics[height=4.8cm]{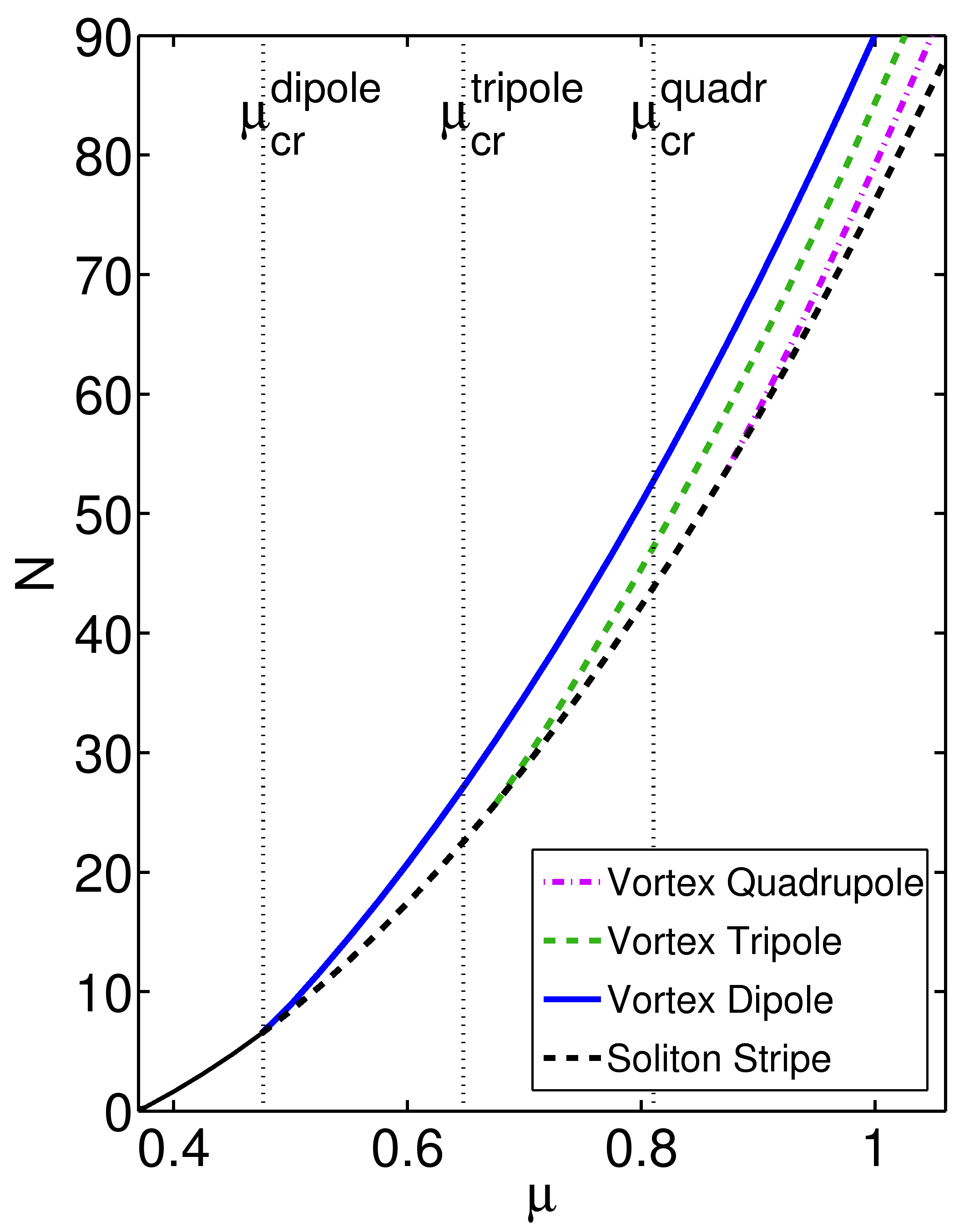}
\label{fig:bif0-7}
}
\caption[Optional caption for list of figures]{\label{fig:bif}Bifurcation diagrams for different values of the anisotropy parameter. 
Stable branches are shown as solid, unstable branches as dashed lines. 
Predictions for the bifurcation points based on the Galerkin-type approach are given as dash-dotted vertical lines.}
\end{figure}

Eq. (\ref{eqncr}) implies that $N_{\text{cr}}$ crucially depends on the energy difference of $\psi_{10}$ and $\psi_{0n}$, which in turn is controlled by the value of the anisotropy parameter $\alpha$.
Choosing $\alpha > 1$, the energy difference is increased, and the bifurcations leading to aligned vortex states occur at higher values of $N$ (or $\mu$) than in the isotropic case.
Physically speaking, the admixture of a $\rm{i} \psi_{0n}$ component to $\psi_{10}$ is energetically suppressed for large $\omega_y$.
As a special case of this, the degeneracy of $\psi_{01}$ and $\psi_{10}$ underlying the single vortex state is lifted, and their superposition does no longer form a stationary state in the linear limit.
Instead, the single vortex bifurcates from the soliton stripe along the $y$-axis at a finite particle number.
This is now the first bifurcation from the solitonic branch, leading to its destabilization.
The dipole then bifurcates from the already destabilized soliton stripe and is thus unstable as well, in full agreement with the results from the particle picture and our numerics.
The single vortex branch, on the other hand, inherits the soliton's stability and is stable for any $\alpha > 1$.
The numerically determined bifurcation diagram for $\alpha=1.5$ is shown in Fig. \ref{fig:bif1-5}, together with the critical values of $\mu$ calculated using the Galerkin approach.

Conversely, for values of $\alpha < 1$ the energy difference in the 
numerator of Eq. (\ref{eqncr}) gets smaller, and the bifurcation points are shifted towards smaller values of $N$ and $\mu$.
Considering the single vortex state first, we find that it no longer bifurcates from the $\psi_{10}$ soliton stripe branch in this anisotropic regime.
This can be expected, as for $\alpha < 1$ the energy of $\psi_{01}$ is lower than that of $\psi_{10}$, and so the single vortex bifurcates from the energetically favourable $\psi_{01}$ soliton stripe branch, characterized by zero density along the $x$-axis.
As a consequence, the single vortex is no longer included in the bifurcation diagram for $\alpha=0.7$, shown in Fig. \ref{fig:bif0-7}.
Comparing to the $\alpha=1.5$ case clearly illustrates how the bifurcations of the dipole and tripole branches are shifted to smaller particle numbers.

Lowering $\alpha$ even further continues this trend, and eventually at 
$\alpha=1/2$ the vortex dipole arises in the linear limit, due to the degeneracy of $\psi_{10}$ and $\psi_{02}$.
For values $\alpha < 1/2$, the dipole bifurcates from the now 
energetically favorable two soliton branch starting from $\psi_{02}$.
Concerning stability, this means that the bifurcation of the dipole can no longer destabilize the soliton stripe branch for $\alpha \leq 1/2$, which in turn implies that the tripole branch bifurcates from the {\it stable} soliton 
stripe branch and inherits its stability in this regime.
This, again, agrees well with the numerically found BdG spectrum in Fig. \ref{fig:bdg_vt}.

The picture keeps repeating for decreasing values of $\alpha$: 
Generally, for $\alpha = 1/(n-1)$, the aligned $(n-1)$ vortex state 
emerges in the linear limit, and the aligned $n$-vortex state takes its role 
in the bifurcation diagram and becomes stable, as it is the first state to bifurcate from the soliton stripe branch.
This shows that the $n$-vortex state is stable in the regime $1/n \leq \alpha \leq 1/(n-1)$. 
Additionally, for $\alpha < 1/n$, the aligned $n$-vortex state is the first state bifurcating from the solitonic branch starting at $\psi_{0n}$, inheriting 
that state's stability.

Thus, in total, we conclude that the state with $n$ vortices aligned along the $y$-axis is stable for any $\alpha \leq 1/(n-1)$.
In particular, aligned vortex states with an arbitrary number of vortices can be stabilized using strong enough transverse confinement. 
Note that this very general result agrees well with our numerical findings 
for the single vortex, vortex dipole and vortex tripole cases discussed above.

\begin{figure}[ht]
\centering
\fbox{
\subfigure{
\includegraphics[width=7cm]{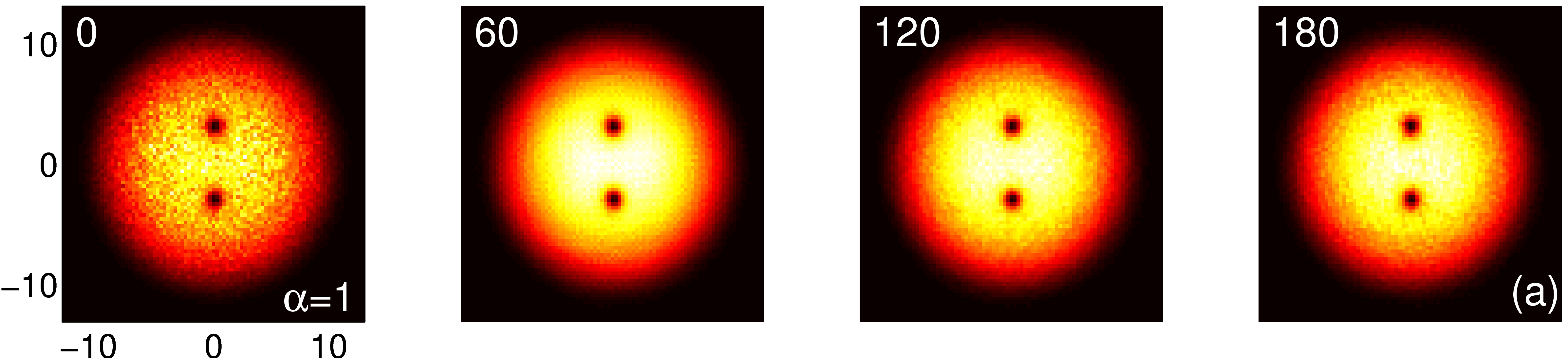}
\label{fig:prop_vd1}
}
}
\fbox{
\subfigure{
\includegraphics[width=7cm]{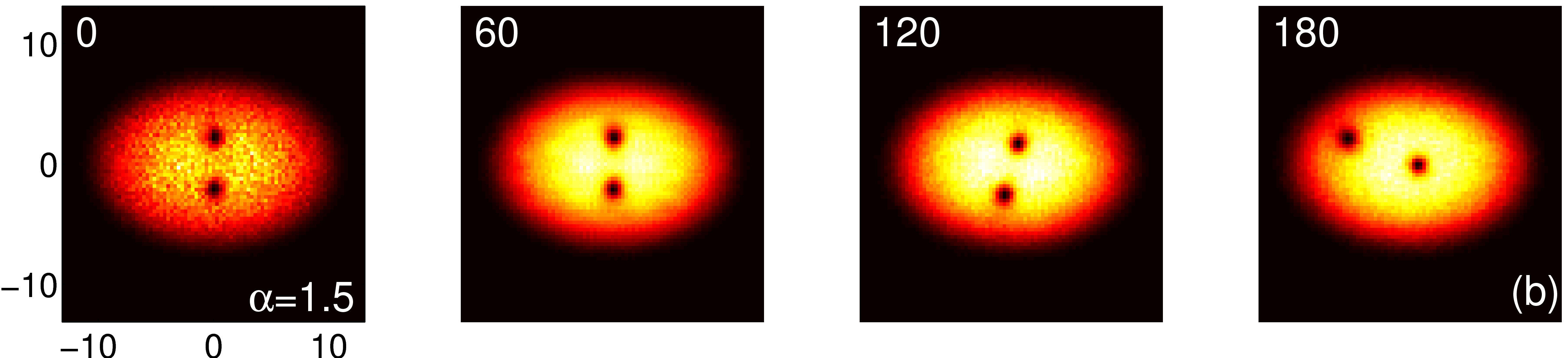}
\label{fig:prop_vd1-5}
}
}
\fbox{
\subfigure{
\includegraphics[width=7cm]{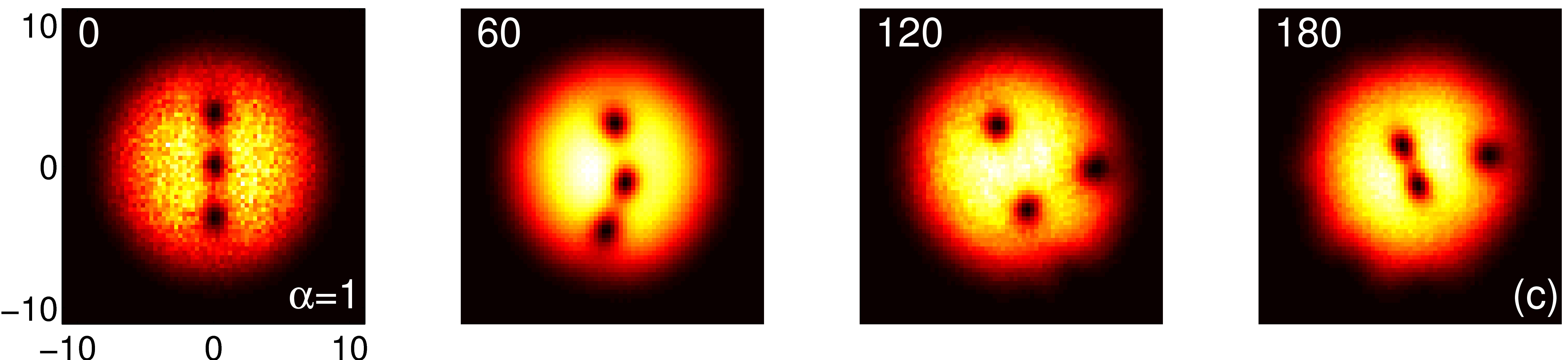}
\label{fig:prop_vt1}
}
}
\fbox{
\subfigure{
\includegraphics[width=7cm]{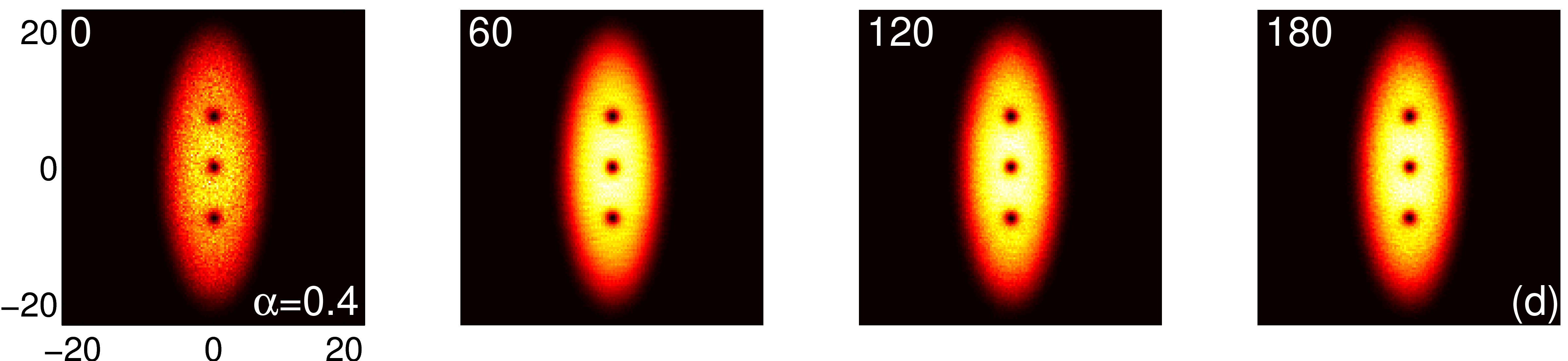}
\label{fig:prop_vt0-4}
}
}
\caption[Optional caption for list of figures]{\label{fig:prop} Time 
propagation of aligned vortex states seeded with white noise for different values of $\alpha$: (a) $\alpha=1$ dipole, (b) $\alpha=1.5$ dipole,
(c) $\alpha=1$ tripole, (d) $\alpha=0.4$ tripole. All plots show $|\psi|^2$, elapsed time is given in the upper left corners, horizontal direction is $x$, vertical $y$.}
\end{figure}

We corroborate the above general conclusions about the role of
anisotropy and its different regimes with numerical simulations 
examining the robustness of the obtained aligned vortex states.
To this end, a white noise signal is added to the state under consideration, 
and it is then propagated in time.
Our observations are in excellent agreement with the stability analysis:
in regimes of $\alpha$ where the vortex states are found to be linearly 
stable, they robustly persist.
On the other hand, for $\alpha$'s where we find imaginary modes in the
BdG spectrum, the perturbation leads far from the equilibrium.
Specifically, four examples are considered in Fig. \ref{fig:prop}.
The top panel confirms the persistence of the vortex
dipole in an isotropic setting. The second row illustrates the destabilization of the same
structure under $\alpha=1.5$ (with the BEC elongating perpendicularly
to the dipole axis) and its transformation into a dynamic
state with one vortex weakly precessing close to the center, while
the other counter-precesses far from it. The tripole dynamics 
is shown 
in the third and fourth rows: while the isotropic
third row case suffers from instability, where again one of the
vortices gets ``expelled'' from the BEC center, the anisotropic
fourth row case of $\alpha=0.4$ restabilizes the tripole in
a BEC elongated along the tripole axis.

\begin{figure}[ht]
\centering
\fbox{
\includegraphics[width=7cm]{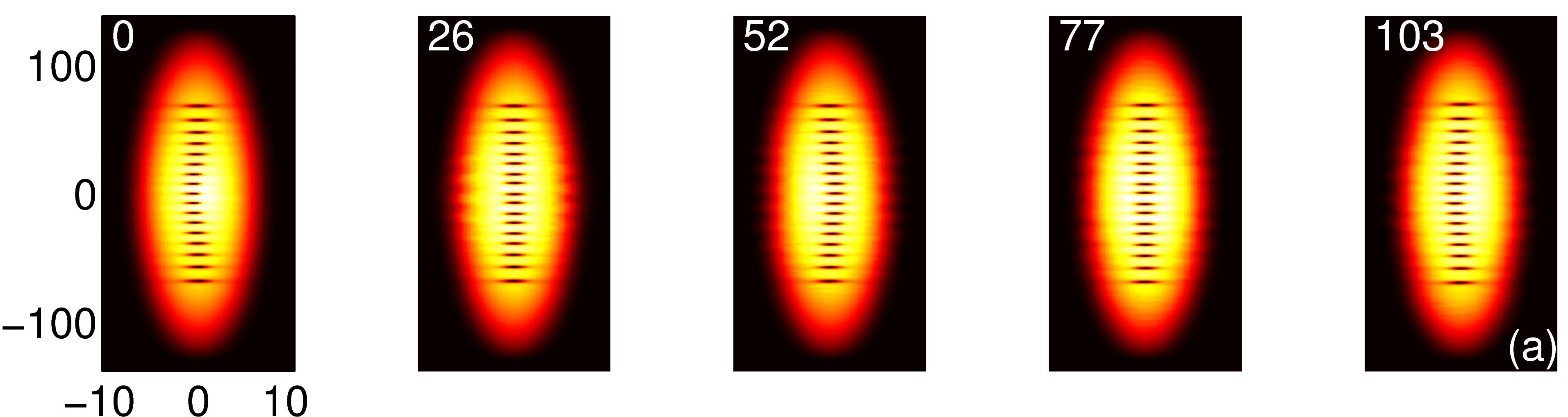}
}
\fbox{
\includegraphics[width=7cm]{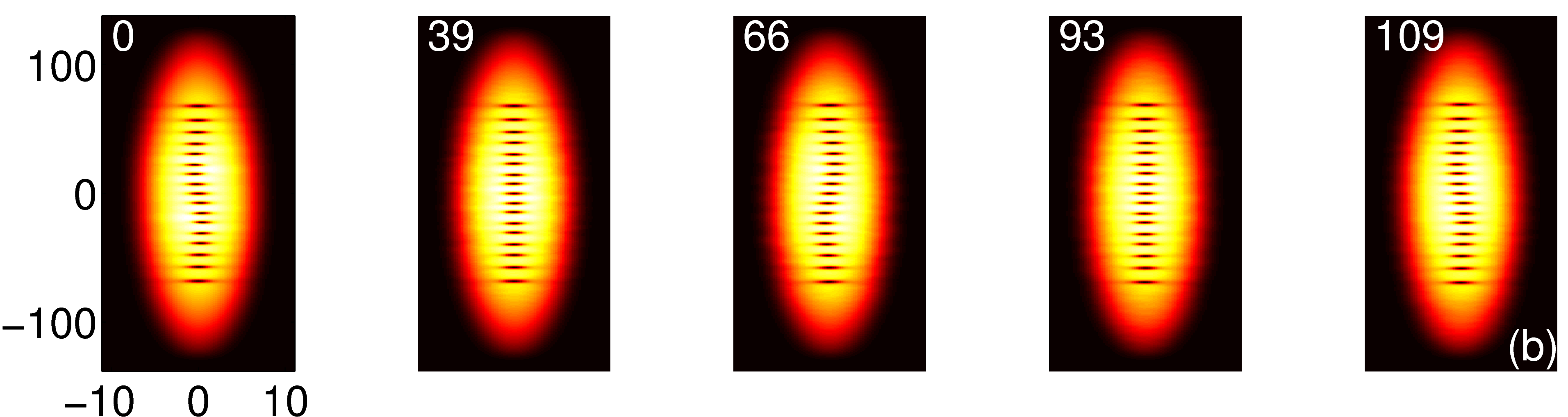}
}
\fbox{
\includegraphics[width=7cm]{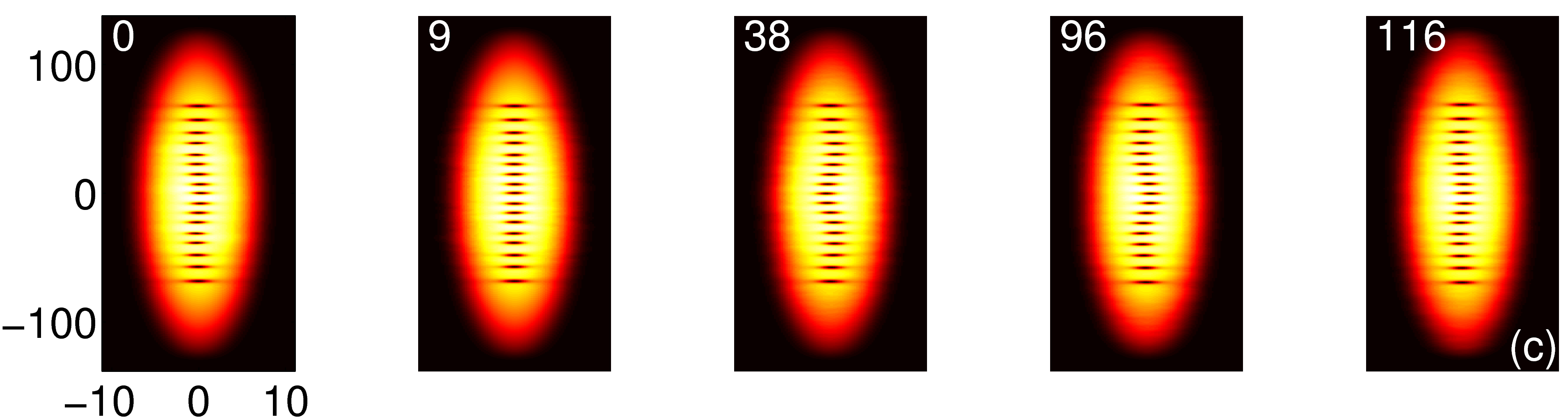}
}
\caption[Optional caption for list of figures]{\label{fig:17vorpics}Time propagation of $|\psi|^2$ of an aligned 17 vortex state when disturbed with eigenvectors belonging to different anomalous modes from the BdG spectrum ($\alpha=0.06$). 
One period of the fundamental vibration (a) at $\omega=0.0609$ and the first two overtones (b,c)
at $0.0574$ and $0.0539$ are shown.
Note the different length scales in the (horizontal) $x$- and (vertical) $y$-direction.}
\end{figure}

Finally, to illustrate that our picture of the aligned vortex states 
and their stabilization is 
valid even for very large numbers of vortices, we discuss a more exotic
state with 17 vortices aligned along the y-axis.
According to the general result reported above, this state should be stabilised 
for $\alpha \leq 1/16 =0.0625$, and indeed we find that for $\alpha=0.06$ it 
remains intact when disturbed with white noise.
Remarkably, exciting the 17 vortex state with eigenvectors belonging to 
anomalous modes from its BdG spectrum leads to a behaviour strongly 
reminiscent of a (discrete variant of a) classical string along the y-axis 
with fixed endpoints, 
supporting a standing wave. Thus, the anisotropy transcending
the dimensional instability barrier allows the observation of a
fundamental vibration mode and the first few ``overtones'' as is
shown in Fig. \ref{fig:17vorpics}. Naturally, the period of such modes is
directly associated with their BdG frequencies of our analysis.

\FloatBarrier

{\it Conclusions.} 
In the present work, we demonstrated how anisotropy 
can be used to manipulate the stability and dynamics of vortex clusters. 
If used to elongate the system in a direction perpendicular
to that of aligned vortex states, it generically destabilizes them.
More importantly, as the system becomes increasingly elongated along the direction of the aligned vortices it can be controllably stabilized below a critical anisotropy and can exhibit
even robust oscillatory dynamics reminiscent of classical strings.
Critical values of the anisotropy for the stabilization of aligned $n$-vortex clusters are provided.
We believe that both our near-linear bifurcation analysis and our particle approach including anisotropic
precessions and vortex interactions can be of value to other fields
where vortex dynamics is of interest, such as nonlinear optics and
condensed matter physics.

\end{document}